\documentclass[aps,prl,floatfix,twocolumn,nofootinbib]{revtex4}
\usepackage{amsmath}
\usepackage{epsfig}
\usepackage{subfigure}
\usepackage{graphicx}
\usepackage{units}
\usepackage{epstopdf}
\long\def\symbolfootnote[#1]#2{\begingroup%
\def\thefootnote{\fnsymbol{footnote}}\footnote[#1]{#2}\endgroup}

\begin{document}
\title{Quantum computing using shortcuts through higher dimensions}
\author{B. P. Lanyon$^{1}$, M. Barbieri$^{1}$, M. P. Almeida$^{1}$, T. Jennewein$^{1,2}$, T. C. Ralph$^{1}$, K. J. Resch$^{1,3}$, G. J. Pryde$^{1,4}$, J.~L.~O'Brien$^{1,5}$, A. Gilchrist$^{1,6}$ \& A. G. White$^{1}$}
\affiliation{$^{1}$Department of Physics and Centre for Quantum Computer Technology, University of Queensland, Brisbane 4072, Australia\\
$^{2}$Institute for Quantum Optics and Quantum Information, Austrian Academy of Sciences, Boltzmanng. 3, A-190 Vienna, Austria\\
$^{3}$Institute for Quantum Computing, University of Waterloo, N2L 3G1, Canada\\
$^{4}$Centre for Quantum Dynamics, Griffith University, Brisbane 4111, Australia\\
$^{5}$Centre for Quantum Photonics, University of Bristol, Bristol BS81UB, UK\\
$^{6}$Physics Department, Macquarie University, Sydney 2109, Australia}

\begin{abstract}

Quantum computation offers the potential to solve fundamental yet otherwise intractable problems across a range of active fields of research. Recently, universal quantum-logic gate sets---the building blocks for a quantum computer---have been demonstrated in several physical architectures. A serious obstacle to a full-scale implementation is the sheer number of these gates required to implement even small quantum algorithms. Here we present and demonstrate a general technique that harnesses higher dimensions of quantum systems to significantly reduce this number, allowing the construction of key quantum circuits with existing technology. We are thereby able to present the first implementation of two key quantum circuits: the three-qubit Toffoli and the two-qubit controlled-unitary. The gates are realised in a linear optical architecture, which would otherwise be absolutely infeasible with current technology.

\end{abstract}

\maketitle

The realisation of a full-scale quantum computer presents one of the most challenging problems facing modern science. Even the implementation of small scale quantum algorithms requires an unprecedented level of control over multiple quantum systems and a deep understanding of quantum mechanics. Recently much progress has been made with demonstrations of universal quantum gates sets - the fundamental building blocks of a quantum computer -  in a number of physical architectures including ion-traps\cite{Schmidt-Kaler:2003qe,Leibfried:2003ai}, linear optics\cite{OBrien:2003lr,Gasp}, and superconductors\cite{NatureSuperC}. In theory these gates can now be put together to implement any quantum algorithm and build a scalable quantum computer. However, in practice there are many significant obstacles that will require much theoretical and technological development to overcome. One of the most daunting is the sheer number of gates required to implement quantum algorithms.

Most approaches to quantum computing employ qubits---the quantum version of bits---encoded in two-level quantum systems. However, candidate systems for encoding quantum information typically have a far more complex physical structure with many readily accessible degrees of freedom, such as atoms\cite{Mandel:2003oq}, ions\cite{Schmidt-Kaler:2003qe,Leibfried:2003ai} or photons\cite{OBrien:2003lr,Gasp}.  In this paper we show how harnessing these higher dimensions of quantum systems during computation can drastically reduce the number of gates required to implement key quantum circuits. This is based on a recent proposal\cite{ralph:022313}, which we extend and use to build two key quantum logic gates for the first time. The technique has the potential for application in a wide range of physical architectures in combination with existing technology,  and offers numerous advantages inherent in a reduced circuit complexity.

One of the most important quantum logic gates that has yet to be realised is the Toffoli---a three-qubit entangling gate that flips the logical state of the `target' qubit conditional on the logical state of the two `control' qubits. Famously, Toffoli gates allow universal reversible classical computation. 
The Toffoli also plays a central role in quantum error correction\cite{PhysRevLett.81.2152} and fault tolerance\cite{PhysRevA.63.052314}.  Furthermore, the combination of the Toffoli and the one-qubit Hadamard offers a simple universal quantum gate set\cite{shi-2002}. The implementation of this gate requires a new level of coherent control over multiple quantum systems and represents a significant experimental challenge. We note that previous demonstrations in a liquid NMR architecture have since been shown to represent, at best, a classical simulation of quantum logic leading to no computational advantage\cite{PhysRevLett.83.1054, PhysRevLett.88.167901}.

The Toffoli-sign (\textsc{ts}) is a three-qubit gate that applies a sign shift to one logical input state and is equivalent to a Toffoli under the action of only a few additional one-qubit gates. The simplest decomposition\cite{MikeIke} of a \textsc{ts} gate when restricted to \textit{qubits} throughout the calculation requires five two-qubit gates. 
If we further restrict ourselves to controlled-\textsc{z} (or \textsc{cnot}) gates this number climbs to six\cite{MikeIke}, Fig.~1a. A quantum logic circuit that requires only three two-qubit gates is shown in Fig.~1b\cite{ralph:022313}. The increased efficiency is achieved by harnessing a third level of one of the information carriers  \emph{during} computation---i.e. one carrier becomes a qutrit with logical states \textbf{0}, \textbf{1} and \textbf{2}. The action of the first $X_a$ gate is to move information encoded in the logical \textbf{0} state of this qutrit into its third level ($\mathbf{2}$), which then bypasses the subsequent two-qubit gates. The final $X_a$ gate \textit{coherently} brings this information back into the \textbf{0} state, reconstructing the logical qubit. This shortcut through a higher dimension allows the two-qubit gates to operate on a subspace of the original target qubit and implementation of the TS with a reduced number of gates. Note that standard two-qubit gates are necessary, with only the additional requirement that they apply the identity to information encoded in the logical $\mathbf{2}$ state of the qutrit. The technique can be generalized to implement higher-order $n$-control-qubit Toffoli gates ($^n\textsc{t}$) by harnessing a single ($n{+}1$)-level information carrier during computation and only ($2n{-}1$) standard two-qubit gates\cite{ralph:022313} (see supplementary information).

We extend this approach to simplify the construction of another key quantum circuit: the two-qubit controlled-unitary (\textsc{cu}) which applies an arbitrary one-qubit gate (\textsc{u}) to a target qubit conditional on the state of a single control qubit, Fig.~1b. This circuit plays a central role in quantum computing, particularly in the phase estimation algorithm\cite{MikeIke} which in turn finds application in quantum simulation 
and quantum chemistry\cite{AlanAspuru-Guzik09092005}. Phase estimation also underpins Shor's famous quantum algorithm for factoring numbers\cite{shor}, the implementation of which would have significant consequence for modern cryptography. Futhermore, the set of \textsc{cu}'s alone is sufficient for universal quantum computing; a \textsc{cu} can implement a \textsc{cnot} and induce any single qubit rotation at the expense of an additional ancilla qubit, for example.  Our technique can be generalized to implement higher-order n-control-bit unitary ($\textsc{c}^n\textsc{u}$) gates using an ($n{+}1$)-level target and only $2n$ two-qubit gates (see supplementary information). Even more generally, the approach can be used to efficiently add $n$ control-qubits to an arbitrary controlled unitary that operates on $k$ qubits (see supplementary information).

The simplest known decompositions\cite{MikeIke} of a $^n\textsc{t}$ and  $\textsc{c}^n\textsc{u}$ into two-qubit gates requires $(12n{-}11)$ and ($12n{-}10$) gates, respectively\symbolfootnote[3]{We note that an architecture-specific decomposition of the $^{n}\textsc{t}$ has been proposed for implementation in an ion trap\cite{PhysRevLett.74.4091}. This requires ($2n{+}1$) qubit-\textit{qutrit} entangling gates.}. In each case this is achieved by employing an additional overhead of ($n{-}1$) ancilla qubits. (Note that implementations without ancillas require on the order of $n^2$ two-qubit gates\cite{MikeIke}). By harnessing only higher dimensions of \emph{existing} information carriers we achieve a significant reduction in all cases. For example, a known decomposition\cite{MikeIke} of the $^5\textsc{t}$ and $\textsc{c}^5\textsc{u}$ both require 50 two-qubit gates plus $4$ ancilla qubits, when restricted to \textit{qubits}. Our technique requires only 9 for the former, 10 for the latter and no ancillary information carriers in either case. 

\begin{figure}
\vspace{0mm}
\includegraphics[width=1 \columnwidth]{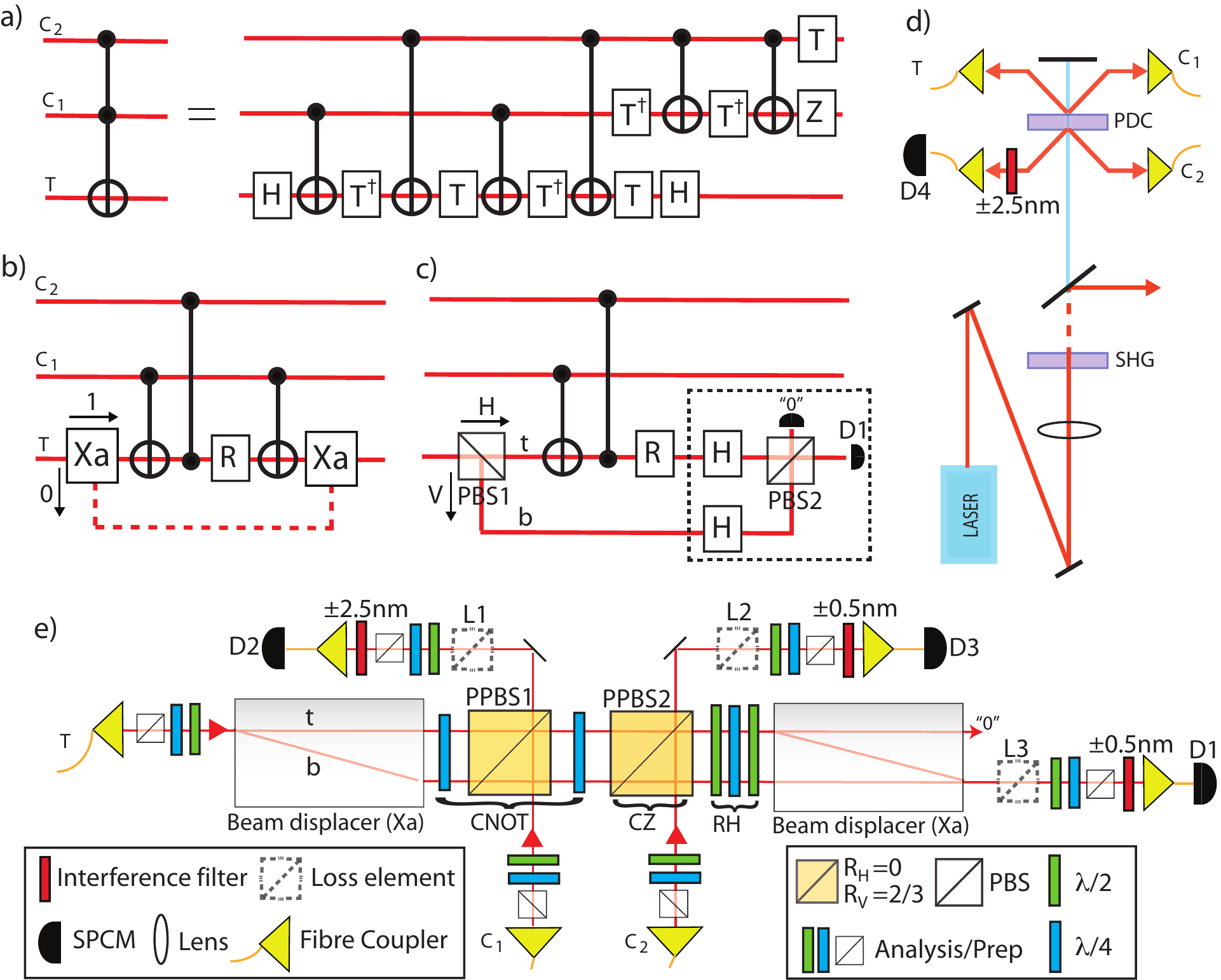}\
\caption{a) Quantum logic circuit for a three-qubit Toffoli gate requiring six two-qubit \textsc{cnot} gates when restricted to employing only qubits\cite{MikeIke}. Standard symbols are used (see supplementary information). b) Circuit for a Toffoli Sign or controlled-$\textsc{z}_{\theta}$ gate. The Toffoli Sign gate ($\textsc{r}{=}\textsc{i}$, the identity) requires only three two-qubit gates (two \textsc{cnot}'s and one \textsc{cz}). A sign change occurs only on the input $|C_2,C_1,T\rangle{=}|\mathbf{1},\mathbf{0},\mathbf{1}\rangle$. For the controlled-unitary $\textsc{r}{{=}}\textsc{z}_{\theta}$ (imparting an arbitrary phase shift between logical states) the circuit implements a controlled-$\textsc{z}_{\theta}$ between $C_1$ and $T$. $C_2$ and the gate it controls are redundant in this case. $X_a$ expands the Hilbert space of the target ($T$) from a \emph{qubit} to a \emph{qutrit}. Its action is defined by: $X_a |\mathbf{0}\rangle{=}|\mathbf{2}\rangle,  X_a |\mathbf{2}\rangle {=} |\mathbf{0}\rangle, X_a |\mathbf{1}\rangle {=} |\mathbf{1}\rangle$. The two-qubit gates act on the qubit levels in the usual way\cite{MikeIke} and always impart the identity on a qutrit in level \textbf{2}. 
c) Conceptual logic circuit for our linear-optic implementation. The final \textsc{cnot} in b) has been replaced by a non-deterministic recombination of the target qubit (dashed box).
d) Layout of our photon source. Forward and backward photons pairs are produced via parametric downconversion (PDC) of a frequency-doubled mode-locked Ti:Saph laser (\unit[820]{nm} ${\rightarrow}$\unit[410]{nm}, $\Delta \tau{{=}}\unit[80]{fs}$ at \unit[82]{MHz} repetition rate) through a Type I PDC \unit[2]{mm} BiB$_{3}$O$_{6}$ crystal. The photons are  collected into four single-mode optical fibres 
and detected using fibre-coupled single photon counting modules. 
e) Experimental implementation of the conceptual circuit shown in c). A successful run of the \textsc{ts} (\textsc{cu}) is flagged by a four-fold (two-fold) coincident measurement at D1-4 (D1-2). 
We employ an inherently stable Jamin-Lebedev interferometer using two calcite beam-displacers\cite{OBrien:2003lr} to coherently expand the dimension of $T$ from a polarisation qubit to a four-level system distributed across polarisation and spatial-mode. We spectrally filter 
using unblocked interference filters centered at \unit[820]{nm} (bandwidths as shown) and realise an arbitrary one-qubit gate (\textsc{rh}) using a quarter-half-quarter waveplate combination.} 
\vspace{-5mm}
\label{fig:algo}
\end{figure}

With an abundance of higher dimensions available in most candidate systems for encoding quantum information, our technique has the potential for wide application. A clear example is the photon, which has a large number of degrees of freedom including polarisation, transverse spatial-mode, arrival-time,  
photon number, and frequency. Coherent control over and between many of these dimensions has already been demonstrated and shown to offer significant advantages in a range of applications such as quantum communication and measurement\cite{kwiatGrover, schuck:190501}. Trapped ions also offer a range of higher dimensions that can be readily exploited, including additional electronic and vibrational energy levels. Indeed, these levels are routinely used for read-out and to coherently implement individual logic gates\cite{Roos:2004fr, PhysRevLett.74.4091}. An immediate benefit of a significant reduction in the number of two-qubit gates required for quantum circuits is an equally significant speed-up in processing time. This has particular advantages for the many architectures where short coherence times are an obstacle in the path to scalability. 

Here we present an implementation using photons to encode information and linear optics to construct quantum gates. Such gates are high performing, well characterised and offer fast gate speeds. Linear optics provide an excellent test-bed for studying quantum information science, and has several known paths to scalable quantum computing\cite{Knill:2001lr, nielsen:040503}. Without the resource saving technique that we present here linear optic implementations of these gates is \emph{absolutely} infeasible with current technology. We note that our resource saving scheme is fundamentally different from and  potentially complementary to schemes for reducing the overhead associated with generating a universal resource\cite{PhysRevLett.91.037903, ralph:100501, browne:010501}; here we are concerned with reducing the amount of that resource required to implement algorithms.

Fig.~1c shows a conceptual circuit of our implementation of the \textsc{ts} and \textsc{cu}. Key steps are the expansion of the Hilbert space of the target qubit ($T$), effected by the first polarising beamsplitter (PBS1), and contraction back into the original space, effected by the components in the dashed box. 
Before PBS1 we have a two-level system in the target rail with logical states $\mathbf{H}{=}\mathbf{0}$ and $\mathbf{V}{=}\mathbf{1}$ (a polarisation qubit). PBS1 then moves information encoded in the logical $\mathbf{H}$ state into a separate spatial mode. After PBS1 we have access to a \textit{four-level} system; two levels in the top rail ($\mathbf{t}$) and two in the bottom rail ($\mathbf{b}$), with logical basis states $|\mathbf{H},\mathbf{t}\rangle$, $|\mathbf{V},\mathbf{t}\rangle$,   $|\mathbf{H},\mathbf{b}\rangle$ and $|\mathbf{V},\mathbf{b}\rangle$, respectively.  While we only need to use one of the additional levels in the bottom rail, we use both in our experiment simply to balance optical path-lengths. The contraction back into the original two-level polarisation qubit is performed non-deterministically, i.e. given deterministic two-qubit gates measurement of a single photon at D1 heralds a successful run of the gate.
This allows for a demonstration without the last \textsc{cnot} gate of Fig.~1b, thereby making an implementation feasible with recent developments in linear optics quantum gates\cite{lanyon-2007}. 
 
 In the case where \textsc{r} is the identity, the circuit in Fig.~1c implements a \textsc{ts} gate between carriers $C_2$, $C_1$ and $T$, applying a sign shift to the $|C_2,C_1,T\rangle{=}|\mathbf{1},\mathbf{0},\mathbf{1}\rangle$ term. It is of no consequence which state receives the sign shift; all cases are equivalent to a Toffoli under the action of additional one-qubit gates. For polarisation encoding, such gates are trivial to implement with standard birefringent waveplates. In the case where \textsc{r} is $\textsc{z}_{\theta}$ the circuit implements a controlled-$\textsc{z}_{\theta}$ ($\textsc{cz}_{\theta}$) between $C_1$ and $T$. $\textsc{z}_{\theta}$ imparts an arbitrary phase between the logical states, i.e.  the operation $\textsc{z}_{\theta}|\mathbf{0}\rangle {\rightarrow} |\mathbf{0}\rangle, \textsc{z}_{\theta }|\mathbf{1}\rangle {\rightarrow} e^{i\theta}|\mathbf{1}\rangle$. It is straightforward to show that such a gate is locally equivalent to a \textsc{cu}, under the action of additional one-qubit gates (see supplementary material).  
  
 \begin{figure}
\includegraphics[width=0.9 \columnwidth]{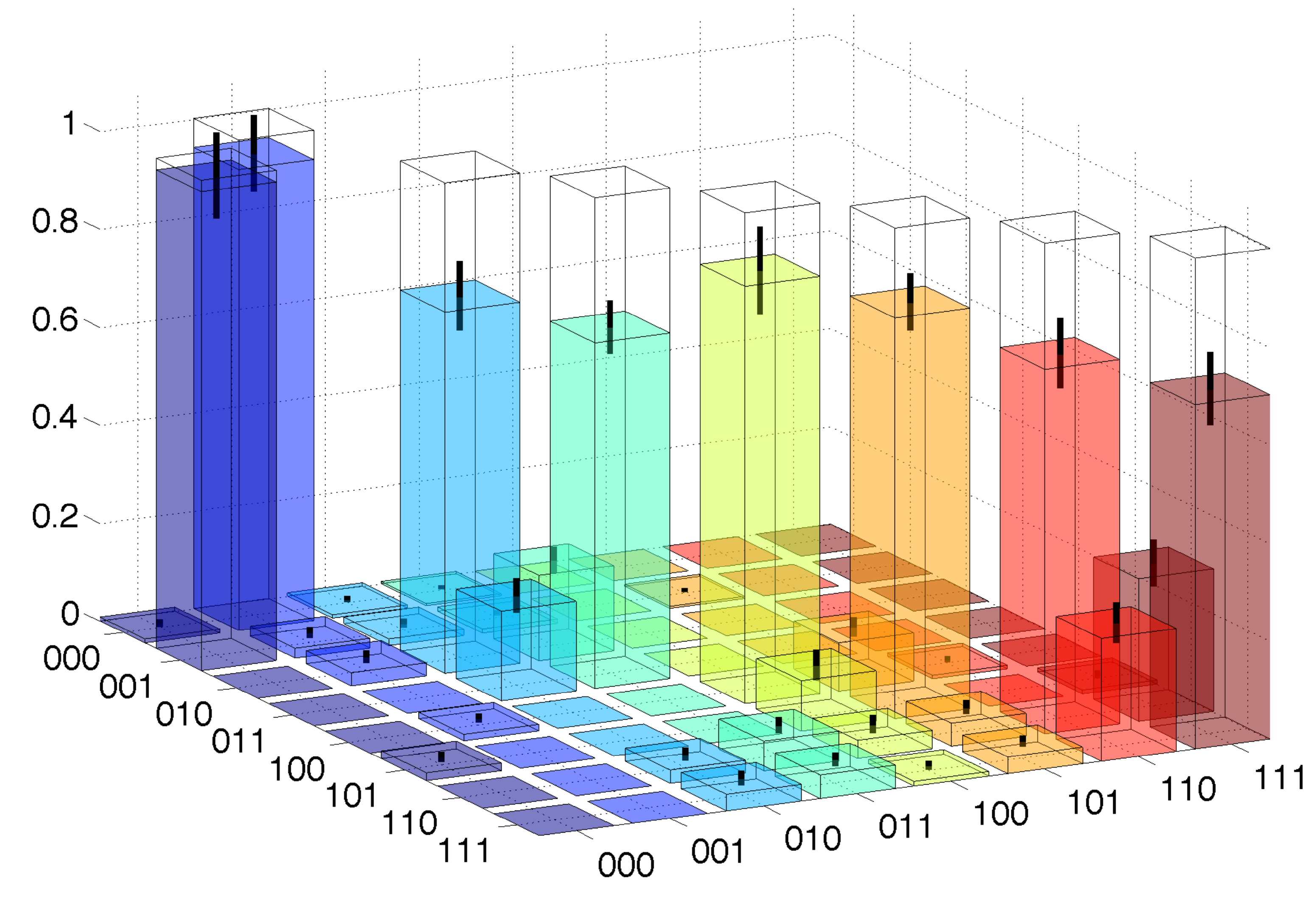}
\caption{Experimentally reconstructed truth-table for our Toffoli gate. Axis labels are written in the order $|C_2,C_1,T\rangle$. Ideally a flip of the logical state of the target qubit ($T$) occurs only when both control qubits ($C_2$ and $C_1$) are in the logical $|\mathbf{0}\rangle$ state. The overlap of ideal and measured truth tables is $I{=}0.81{\pm}{0.03}$. The ideal case is shown as a wire grid. Error bars are shown representing one standard deviation, calculated from Poissonian photon counting statistics. The Inquisition ($\mathcal{I}$) is defined as the average logical state fidelity of a truth table $\bar{\mathcal{I}}{=}\textsc{t}r(\textsc{m}_{\rm{exp}}\textsc{m}_{\rm{ideal}})/d$, where $\textsc{m}_{\rm{exp}}$ and $\textsc{m}_{\rm{ideal}}$ are the measured and ideal truth tables, and $d$ is the table dimension.  We define the logical flipping contrast as $\mathcal{C}{=}1/2\{1+(\textsc{p}_{\rm{ideal}}-\textsc{p}_{\rm{flip}})/(\textsc{p}_{\rm{ideal}}+\textsc{p}_{\rm{flip}})\}$ 
where $\textsc{p}_{\rm{ideal}}$ is the probability of obtaining the ideal output state and $\textsc{p}_{\rm{flip}}$ is the probability of obtaining the output state where the ideal target qubit output state has been flipped.}
\vspace{-5mm}
\label{fig:truth}
\end{figure} 
 
Figs.~1d-e show the layout of our optical source and logic circuit in the laboratory, respectively. Our photons are generated via spontaneous parametric down conversion (SPDC), spatially filtered via single-mode optical fiber and spectrally filtered using interference filters, as detailed in the figure caption. Two-qubit linear optic gates are realised by combining one-qubit gates (waveplates) and controlled-\textsc{z} ($\textsc{cz}{=}\textsc{cz}_{\pi}$) gates based on non-classical interference at partially-polarising beamsplitters; the gates are nondeterministic and employ a measurement induced non-linearity\cite{langford:210504, kiesel:210505, okamoto:210506}.  Rather than chaining the gates to implement the required sequence for our implementation of the Toffoli (Fig.~1c) we employed a recently developed three-qubit quantum logic gate\cite{lanyon-2007}. Ideal operation of the two-qubit gates requires non-classical interference between indistinguishable photons at the PPBS's (Fig.~1e). Dependent photons generated from the first pass of the PDC source interfere non-classically at PPBS1. One photon then goes on to interfere with a third, independent photon from the second pass, at PPBS2. We measure relative non-classical two-photon interference visibilities between vertically polarised photons of $V_{r}{{=}}$100$\pm$1\% and $V_{r}{{=}}$92$\pm$4\% (where $V_{r}{{=}}V_{\mathrm{meas}}/V_{\mathrm{ideal}}$ and $V_{ideal}{=}80\%$), for the dependent and independent interferences respectively. The difference in these visibilities reflects the difficulty of achieving perfectly indistinguishable photons generated from independent sources\cite{rarity1996}. For our experimental implementation we include the additional one-qubit gates required to convert the \textsc{ts} in Fig.~1c to the Toffoli.  In both implementations our imperfectly manufactured beamsplitters impart systematic unitary operations on the optical modes. For simplicity we corrected for these effects numerically. Alternatively such unitaries could be corrected with standard waveplates.

\begin{figure}
\vspace{10mm}
\begin{center}
\includegraphics[width=1 \columnwidth]{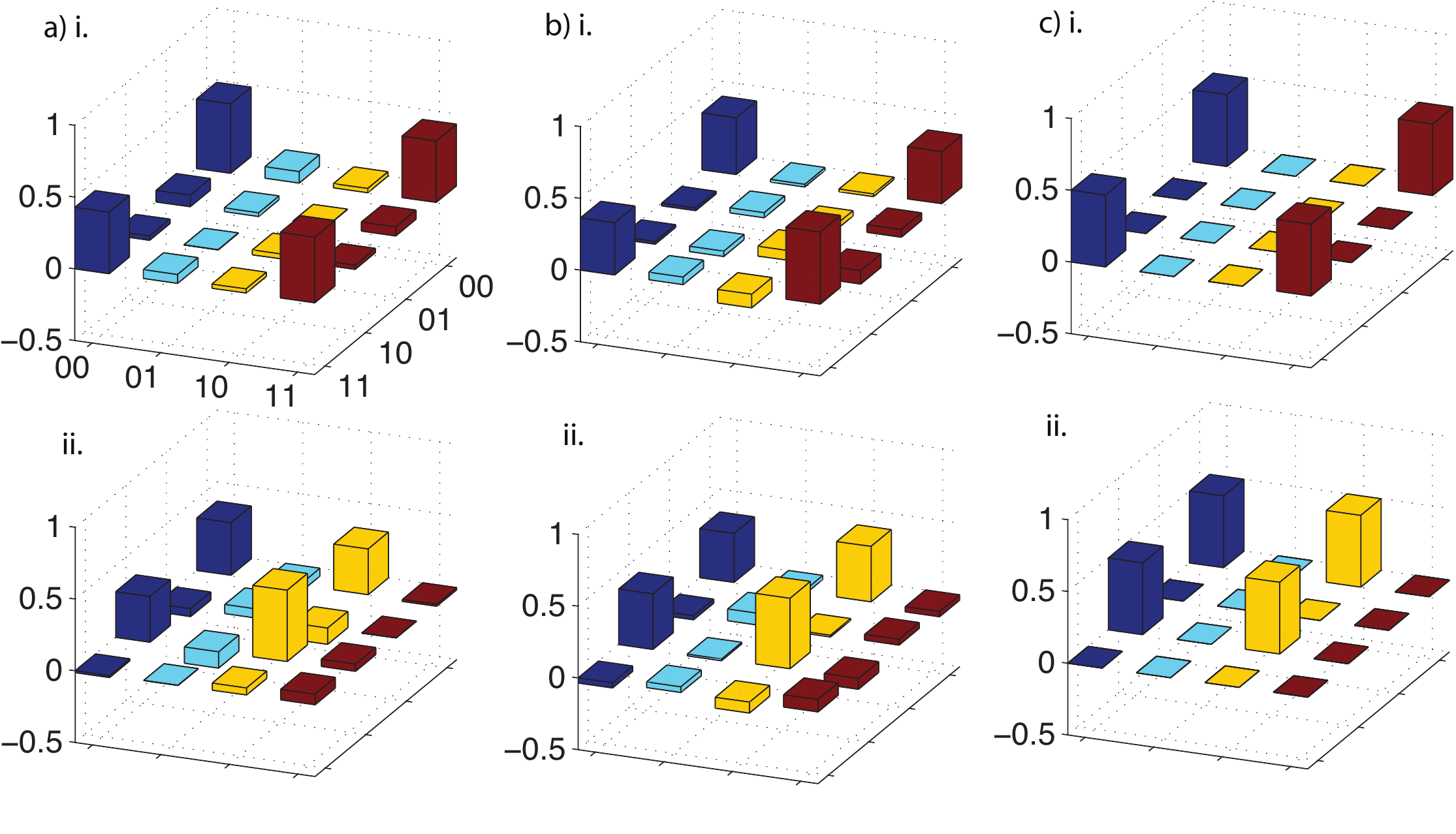}
\end{center}
\caption{a) Measured output states of qubits $C_1$ and $T$ for Toffoli gate inputs; i) $|\mathbf{0},(\mathbf{0}{+}\mathbf{1}),\mathbf{0}\rangle/\sqrt{2}$; and ii) $|\mathbf{1},(\mathbf{0}{+}\mathbf{1}),\mathbf{0}\rangle/\sqrt{2}$. We observe fidelities with the ideal states, linear entropies and tangles\cite{White:07} 
of i) \{$0.90{\pm}{0.04}$, $0.21{\pm}{0.08}$, $0.68{\pm}{0.10}$\} and ii) \{$0.75{\pm}{0.06}$, $0.47{\pm}{10}$, $0.04{\pm}{0.06}$\}, respectively. b) As for a) but where the roles of $C_1$ and $C_2$ have been swapped. We now observe i) \{$0.81{\pm}{0.02}$, $0.39{\pm}{0.05}$, $0.53{\pm}{0.07}$\} and ii) \{$0.80{\pm}{0.03}$, $0.40{\pm}{0.05}$, $0.01{\pm}{0.01}$\}. The decrease in tangle in the i) cases reflects the difference between dependent and independent photon interference, as discussed in the text. c) ideal density matrices. Note in all cases only real parts are shown, imaginary parts are negligible.  The fidelity between two matrices (either two states or two processes) is $\textsc{f}(\rho,\sigma){\equiv}\{\mathrm{\textsc{t}r}\sqrt{\sqrt{\rho} \sigma \sqrt{\rho}}\}^{2}$; Linear entropy is $\textsc{s}_{\textsc{l}} {\equiv}$ $d (1{-}\mathrm{\textsc{t}r}[\rho^2])/(d{-}1)$, where $d$ is the state dimension.}
\label{fig:bell}
\end{figure}

For our implementation of a Toffoli we use photons from both passes of the PDC crystal and set \textsc{r} to the identity (Fig.~1e). A four-fold coincident measurement at detectors D1-4 signals a successful run of the gate. In linear optics implementations of two-qubit quantum gates, state dependent loss is used to rebalance amplitudes\cite{langford:210504, kiesel:210505, okamoto:210506}. When incorporating loss elements L1-3 the gate operates with a success probability of 1/72. Alternatively, to combat low count rates, correct balance can be achieved by removing additional loss elements and pre-biasing the input polarisation states during gate characterisation\cite{langford:210504, kiesel:210505, okamoto:210506}. Under these conditions we measure a four-fold coincidence rate of approximately \unit[10]{mHz} when running at full pump laser power. While this is not sufficient to perform a full process tomography\cite{obrien:080502} of the gate over a practical time period, we are able to to demonstrate all the key aspects of its behavior.  

To demonstrate the classical action of our Toffoli we first reconstruct a logical truth-table (Fig.~2). In the ideal case of our implementation the target ($T$) undergoes a logical flip if, and only if, both control qubits are in the logical $\mathbf{0}$ state. 
The inquisition, $\mathcal{I}$, is the overlap between the ideal and measured truth tables\cite{White:07}: for our three-qubit gate we measure $\mathcal{I}{=}0.81{\pm}{0.03}$, compared to $\mathcal{I}{=}0.85{\pm}{0.01}$ achieved for optical implementations of two-qubit gates\cite{okamoto:210506}. Significant deviations from the ideal correspond to unwanted flips of the target qubit which can be understood with reference to the non-classical interferences required for correct operation in each case.  In order to gauge our gates ability to apply the correct operation to a subset of logical input states we employ the flipping contrast, $\mathcal{C}$. (See Fig.~2). 
For inputs $|C_2,C_1\rangle{=}|\mathbf{0},\mathbf{0}\rangle$, no non-classical interference is required for correct operation and we measure $\mathcal{C}{=}0.99{\pm}{0.01}$, averaged over both target logical input states. Inputs $|C_2,C_1\rangle{=}|\mathbf{0},\mathbf{1}\rangle$ require perfect non-classical interference between \textit{dependent} photons $C_1$ and $T$, for ideal operation. As discussed previously we achieve a near perfect interference visibility between vertical photons in this case. However, the full process suffers from the presence of higher-order PDC terms which are not observed in the visibility measurement due to higher-order interference processes\cite{Tills}.  This is reflected in an average of $\mathcal{C}{=}0.95{\pm}{0.02}$. Inputs $|C_2,C_1\rangle{=}|\mathbf{1},\mathbf{0}\rangle$ require perfect non-classical interference between \textit{independent} photons $C_2$ and $T$, for ideal operation. This process is limited not only by higher-order terms but by inherent distinguishability of these photons\cite{rarity1996}, reflected in an average of $\mathcal{C}{=}0.80{\pm}{0.02}$. Inputs $|C_2,C_1\rangle{=}|\mathbf{1},\mathbf{1}\rangle$ require perfect non-classical interference between both dependent and independent photons, and are therefore the most challenging cases. Here we observe an average of $\mathcal{C}{=}0.73{\pm}{0.05}$. 

To demonstrate the quantum action of our Toffoli we test its ability to \emph{coherently} control an entangling process. With an input state of $|\mathbf{0},(\mathbf{0}{+}\mathbf{1}),\mathbf{0}\rangle/\sqrt{2}$ our Toffoli will produce the entangled state $|\mathbf{0},\Psi_{+}\rangle$, where $|\Psi_{+}\rangle$ is the maximally entangled Bell state\cite{MikeIke} $(|\mathbf{0},\mathbf{0}\rangle {+}|\mathbf{1},\mathbf{1}\rangle)/\sqrt{2}$. With an input state of $|C_2,C_1,T\rangle{=}|\mathbf{1},(\mathbf{0}{+}\mathbf{1}),\mathbf{0}\rangle/\sqrt{2}$, it will produce the separable output state $|\mathbf{1},(\mathbf{0}{+}\mathbf{1}),\mathbf{0}\rangle/\sqrt{2}$.  In the former (latter) case the entangling operation between $C_1$ and $T$ is coherently turned on (off) by $C_2$. We then swap the roles of the control qubits and repeat the test. These processes require coherent interaction between all three qubits. We perform over-complete full state tomography to reconstruct the density matrix of two-qubit output states, requiring 36 separate measurements\cite{langford:210504}. Our tomography employs convex optimisation and Monte-Carlo simulation for error analysis\cite{lanyon-2007}. Measurements sets are taken iteratively, whereby multiple sets---each taking around \unit[1]{hr} to complete---are recorded. This reduces the effect of optical source power fluctuations.

Fig.~3 shows experimentally reconstructed density matrices representing the state of a control and target qubit, at the output of our Toffoli gate. We achieve a high fidelity\cite{White:07} with the ideal states in all cases, as detailed in the figure caption. The entangling process required to achieve Fig.~3a)i relies on interference between dependent photons. The process required to achieve Fig.~3b)i, relies on both dependent and independent interference. This leads to the reduced fidelity and increased linear entropy\cite{White:07} observed in the latter case. These processes demonstrate the coherent action of our Toffoli gate.

Higher order PDC terms are caused when more than one pair of photons is created simultaneously in a single PDC pass (Fig.~1d). It is straightforward to show that the ratio of double to single photon-pair emission is proportional to the pump power.  Thus reducing the power by a factor of 4 should in turn reduce these higher-order contributions from each pass by a factor of 4. Under these conditions we observe a four-fold rate at the output of the Toffoli gate of only $\sim$\unit[0.1]{mHz} and repeat measurement of the average flipping contrast for the most challenging logical inputs $|C_2,C_1\rangle{=}|\mathbf{1},\mathbf{1}\rangle$. We observe a clear improvement from $\mathcal{C}{=}0.73{\pm}{0.05}$ to $\mathcal{C}{=}0.83{\pm}{0.04}$.

For our implementation of the \textsc{cu} circuit we inject only dependent qubits $C_1$ and $T$ into the circuit, from the forward pass through the PDC crystal (Figs~1d-e). When incorporating loss element L1 (Fig.~1e) the gate operates with a success probability of 1/18. For our demonstration this was optimised by removing L1 and pre-biasing only the $C_1$ input state\cite{langford:210504, kiesel:210505, okamoto:210506}. A two-fold coincident measurement at detectors D1-2 signals a successful run of the gate.  When running at 1/4 power we observe a two-fold rate at the output of our circuit of around \unit[100]{Hz}. This is more than sufficient to perform a full process tomography\cite{obrien:080502} of the \textsc{cu} on a timescale of the order of  \unit[2]{hrs}, whilst retaining the advantages of a lower power. 

\begin{figure}
\begin{center}
\includegraphics[width=0.95 \columnwidth]{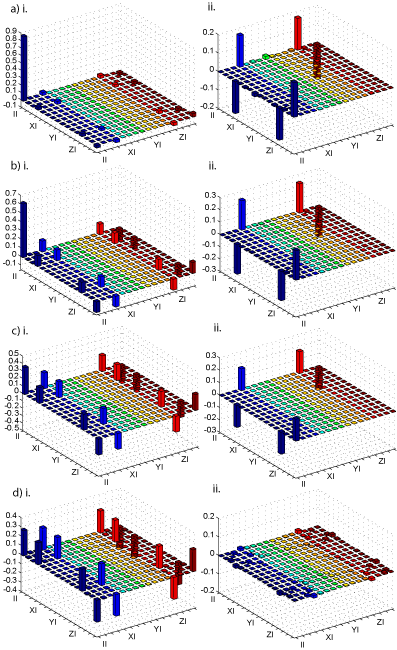}
\end{center}
\caption{Experimentally reconstructed process matrices\cite{obrien:080502} for four demonstrations of our controlled-unitary gate ($\textsc{cu}$), for $\textsc{u}{=}Z_{\theta}$ and: a) $\theta{=}{\pi/4}$ (\textsc{ct}); b) $\theta{=}\pi/2$, (\textsc{cj}); c) $\theta{=}3\pi/4$, (\textsc{cl}); and d) $\theta{=}\pi$, (\textsc{cz}).
(i) Real and (ii) imaginary parts are shown. We observe high process fidelities\cite{obrien:080502} with the ideal 
\{$0.982{\pm}{0.003}$, $0.977{\pm}{0.004}$, $0.940{\pm}{0.006}$, $0.956{\pm}{0.003}$\} and low average output-state linear entropies \{$0.036{\pm}{0.004}$, 
$0.047{\pm}{0.004}$, $0.091{\pm}{0.005}$, $0.086{\pm}{0.006}$\}, respectively. Matrices are presented in the standard Pauli 
basis\cite{obrien:080502}.}
\label{fig:cu}
\end{figure}

As a demonstration, we report the implementation of four distinct \textsc{cu} gates that apply $\textsc{z}_{\theta}$ rotations of $\pi/4$ (\textsc{ct}), $\pi/2$ (\textsc{cj}), $3\pi/4$ (\textsc{cl}) and $\pi$ (\textsc{cz}) to the target ($T$) conditional on the control ($C_1$), respectively. Three of these are of fundamental importance to quantum computing\cite{MikeIke}. We fully characterise these gates via quantum process tomography\cite{obrien:080502}: Fig.~4 shows the experimentally reconstructed process matrices. We achieve the highest reported two-qubit logic gate process fidelities, in any architecture, as detailed in the figure caption. The origins of the small deviations from ideal operation are thought to lie in residual higher order PDC emissions, imperfect mode matching and manufactured optics\cite{rarity1996, Tills}. The overriding source of error in our experiments lies in our imperfect photon source. Current developments in source technology promise significant improvements in the near future. The combination of this with recently developed photon-number resolving detectors offers a path to a deterministic and scalable implementation of our gates.

We have demonstrated a new technique that harnesses the power of readily available higher dimensional quantum systems to drastically simplify the construction of key quantum circuits.  As we have shown, the technique is particularly relevant to the current state of experimental quantum computing and can be integrated with \emph{existing} technology to realise new quantum circuits. The demonstration of these circuits represents a crucial test of the practicality of quantum computing, raises many important questions that have yet to be solved and immediately offers valuable new tools with which to study quantum interactions.

We acknowledge discussions with William Munro and David Kielpinski, and financial support from the Australian Research Council Discovery and Federation Fellow programs, the DEST Endeavour Europe and International Linkage programs, and an IARPA-funded U.S. Army Research Office Contract. 
\\

\vspace{-6 mm}

\

\noindent \textbf{Supplementary material}.

\noindent Our technique can be extended to multiple-control gates, e.g. $^n\textsc{t}$ and $\textsc{c}^n\textsc{u}$. Figure 5 gives an implementation of $^3\textsc{t}$ and $\textsc{c}^2\textsc{u}$. Fig.~6 shows a more general application of our technique; adding $n$ control qubits to an arbitrary $k$ qubit unitary. 

\begin{figure}
\vspace{3mm}
\begin{center}
\includegraphics[width=1\columnwidth]{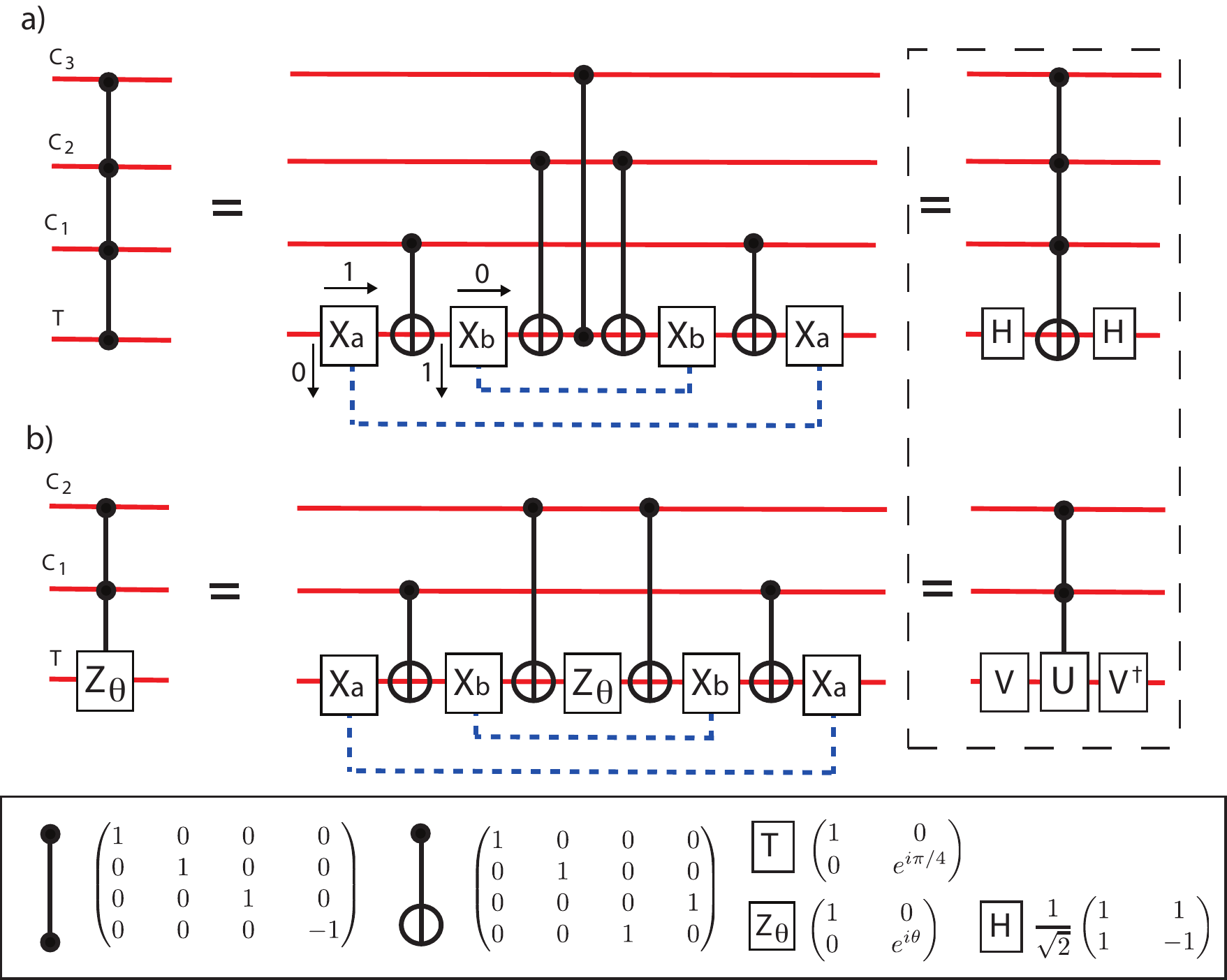}
\end{center}
\vspace{0mm}
\caption{Implementation of the next level of encoding of our technique required to realise \textbf{a} a 3-control-qubit Toffoli Sign gate ($^3\textsc{ts}$) and \textbf{b} 2-control-qubit unitary $\textsc{c}^2\textsc{z}_{\theta}$. The $X_a$ gate flips information between the logical $\mathbf{0}$ and $\mathbf{2}$ state of the target. The $X_b$ gate flips infomation between the logical $\mathbf{1}$ and $\mathbf{3}$ state of the target. Thus we require access to a four-level information carrier in the target; two levels in the original red rail and one in each of the dashed-blue rails. The dashed box shows the 1-qubit gates required to interconvert the $\textsc{ts}^3$ and $\textsc{c}^2\textsc{u}$ to the Toffoli and controlled-unitary equivalents, respectively. $VZ_{\theta}V^{\dagger}$ is the spectral decomposition of U, up to a trivial global phase factor. All other gates operate as usual (on the qubit subspace: levels $\mathbf{0}$ and $\mathbf{1}$) and their operations are shown, including some used in Fig.~1 of the main text. In \textbf{a} the control operation occurs if $|C_3,C_2,C_1\rangle{=}|\mathbf{1},\mathbf{1},\mathbf{0}\rangle$. In \textbf{b} the control operation occurs if $|C_2,C_1\rangle{=}|\mathbf{1},\mathbf{0}\rangle$. a) is based on Fig.~2 of Ref.~3 in the main text.}
\vspace{0mm}
\label{fig1:algo}
\end{figure}

\begin{figure}
\vspace{-3mm}
\begin{center}
\includegraphics[width=0.9\columnwidth]{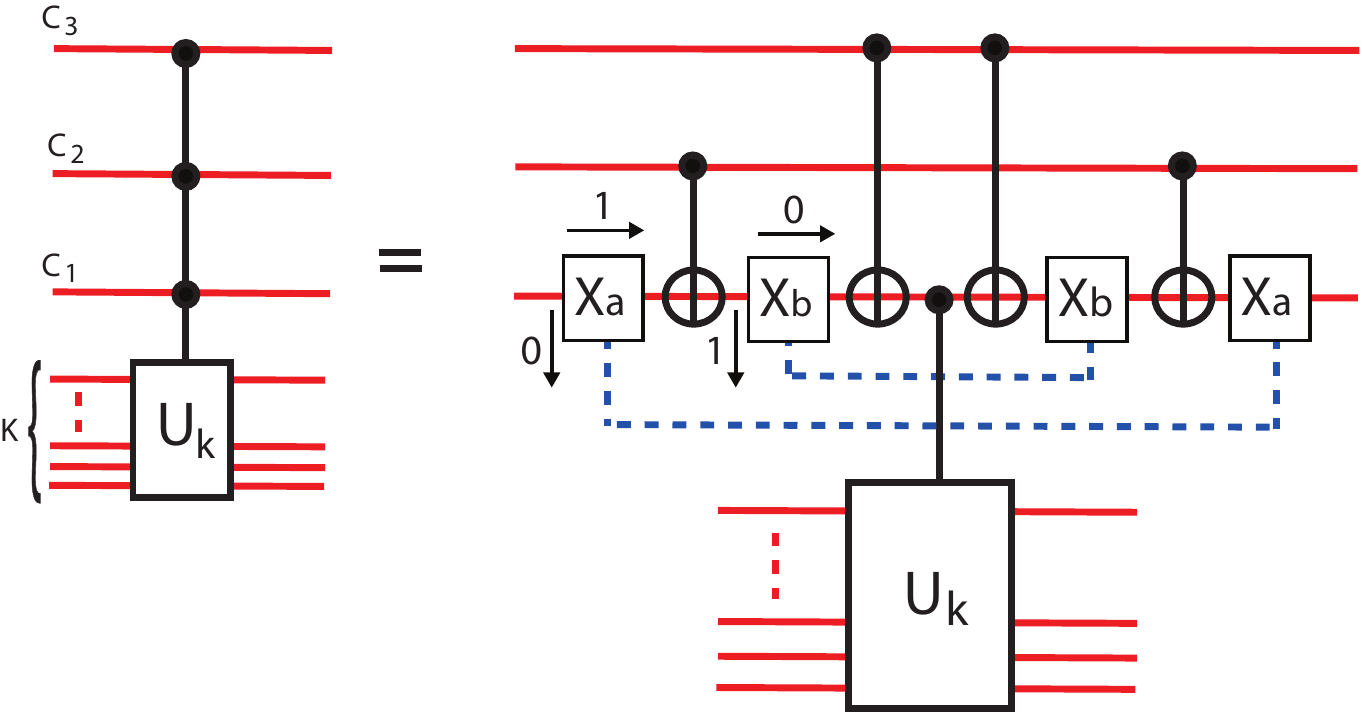}
\end{center}
\vspace{0mm}
\caption{Circuit for implementation of a 3-control-qubit unitary acting on $k$ qubits, $\textsc{c}^3(\textsc{u}_k)$. Given the ability to perform a single instance of a $\textsc{c}^1(\textsc{u}_k)$ $n$ additional control  qubits can be added at a cost of an extra $2n$ two-qubit gates and an additional n dimensions in $C_1$. The $X_j$ perform as described in the caption of Fig.~1.} 
\vspace{0mm}
\label{fig1:algo}
\end{figure}

\noindent \textbf{Correspondence}. 

\noindent Correspondence and requests for materials should be addressed to BPL (lanyon@physics.uq.edu.au). 

\end{document}